\documentclass[twocolumn,aps,epsfig,nofootinbib]{revtex4}

\usepackage{graphicx}
\usepackage{amsfonts}
\usepackage{epstopdf}
\usepackage{latexsym}
\usepackage{amssymb}
\usepackage{amssymb}

\usepackage[center]{subfigure}

\begin{document}

 \newcommand{\bq}{\begin{equation}}
 \newcommand{\eq}{\end{equation}}
 \newcommand{\bqn}{\begin{eqnarray}}
 \newcommand{\eqn}{\end{eqnarray}}
 \newcommand{\nb}{\nonumber}
 \newcommand{\lb}{\label}
\newcommand{\PRL}{Phys. Rev. Lett.}
\newcommand{\PL}{Phys. Lett.}
\newcommand{\PR}{Phys. Rev.}
\newcommand{\CQG}{Class. Quantum Grav.}

\title{On ``No-go theorem for slowly rotating black holes in    Ho\v{r}ava-Lifshitz  gravity"}

 \author{Anzhong Wang}
\email{anzhong_wang@baylor.edu}

\affiliation{ 
GCAP-CASPER, Physics Department, Baylor
University, Waco, TX 76798-7316, USA}

\date{\today}

\begin{abstract}
Slowly rotating black holes in the non-projectable  Ho\v{r}ava-Lifshitz (HL) theory were studied recently in  Phys. Rev. 
Lett. {\bf 109}, 181101 (2012), and claimed that they  do not exist. In this Comment, we show that this is incorrect, 
and  such solutions indeed  exist in the IR limit of the  non-projectable HL theory.
   
\end{abstract}


\maketitle

\section{Introduction}

Recently, Barausse and Sotiriou \cite{BS} studied slowly rotating black holes in the non-projectable version \cite{BPS} of the Horava-Lifshitz (HL) 
theory \cite{Horava}, by using the equivalence  \cite{Jacobson} (See also \cite{BPS}) between the  IR limit of the non-projectable  HL theory  and 
the Einstein-aether 
($\ae$-) theory with the hypersurface-orthogonal condition, and claimed that such  black holes do not exist. This is a very strong statement, and 
immediately causes cautions on the variability of   this version of the HL theory,   because observations  indicate that such black holes very likely  
exist in our universe \cite{Nara}.

In this Comment, we show that the above claim is incorrect, as it was based on three wrong field equations.  After correcting these errors, we find
that slowly rotating black holes indeed exist in the infrared (IR) limit of the  non-projectable HL theory \cite{BPS}.

To this purpose, let us start with the hypersurface-orthogonal $\ae$-theory (For detail, see   \cite{Jacobson}).  
The fundamental variables of the gravitational 
sector in  the $\ae$-theory are $\left(g_{\mu\nu}, u^{\lambda}\right)$, where Greek letters run from 0 to 3, $g_{\mu\nu}$ is  the four-dimensional metric  of the space-time
with the signatures $(+, -,-,-)$,
and  $u^{\lambda}$ the aether four-velocity. The general action of the theory
 is given by,
$S = S_{\ae} + S_{m}$,
where  $S_{m}$ denotes the action of matter,  and $S_{\ae}$  the gravitational action of the $\ae$-theory, given by
\bqn
\lb{1.1} 
S_{\ae} &=& \zeta^2 \int{\sqrt{- g} \; d^4x \Big[- R(g_{\mu\nu}) + {\cal{L}}_{\ae}\left(g_{\mu\nu}, u^{\lambda}\right)\Big]},\nb\\
S_{m} &=& \int{\sqrt{- g} \; d^4x \Big[{\cal{L}}_{m}\left(g_{\mu\nu}, \psi\right)\Big]}.
\eqn
Here $\zeta^2 = 1/(16\pi G_{\ae})$, $\psi$ collectively denotes the matter fields, $R$    and $g$ are, respectively, the  Ricci scalar and determinant of $g_{\mu\nu}$, 
 and 
\bq
\lb{1.2}
{\cal{L}}_{\ae}  = - M^{\alpha\beta}_{~~~~\mu\nu}\left(D_{\alpha}u^{\mu}\right) \left(D_{\beta}u^{\nu}\right),
\eq
 where $D_{\mu}$ denotes the covariant derivative with respect to $g_{\mu\nu}$, and  $M^{\alpha\beta}_{~~~~\mu\nu}$ is defined as
\bq
\lb{1.3}
M^{\alpha\beta}_{~~~~\mu\nu} = c_1 g^{\alpha\beta} g_{\mu\nu} + c_2 \delta^{\alpha}_{\mu}\delta^{\beta}_{\nu}
+  c_3 \delta^{\alpha}_{\nu}\delta^{\beta}_{\mu} + c_4 u^{\alpha}u^{\beta} g_{\mu\nu}.
\eq
Note that here we assume that matter fields couple only to $g_{\mu\nu}$, so ${\cal{L}}_{m}$ is independent of $u^{\mu}$. 
The four coupling constants $c_i$ are all dimensionless, and $G_{\ae}$ is related to  the Newtonian constant $G_{N}$ via the relation,
$G_{N} = {2G_{\ae}}/{(2 - c_{14})}$,
with $c_{14} \equiv c_1 + c_4$, etc. The  hypersurface-orthogonal condition,
\bq
\lb{1.10}
\omega_{\mu} \equiv \epsilon_{\mu\nu\alpha\beta}u^{\nu}D^{\alpha}u^{\beta} = 0,
\eq
implies that there exists a time-like scalar function $T$, so that $u_{\mu}$ is given by   \cite{Wald},
\bq
\lb{1.10a}
u_{\mu} = {T_{,\mu}}/{\left|g^{\alpha\beta}T_{,\alpha} T_{,\beta}\right|^{1/2}},
\eq
where the leaves of constant $T $  naturally provides the foliations  constructed in the HL theory \cite{Horava}.   Then, it is very convenient 
to choose $T$ as the time-like coordinate, and with  
 the ADM  decompositions \cite{ADM},
the aether four-velocity can be expressed as, 
\bq
\lb{1.10b}
u_{\mu} = N \delta^{T}_{\mu},\;\;\; u^{\mu} = \frac{1}{N}\left(\delta_{T}^{\mu} - N^i \delta_{i}^{\mu}\right),\; (i, j  = 1, 2, 3)
\eq
where   $N_{i} = h_{ij}N^j,\; h_{ij} h^{ik} = \delta^{k}_{j}$, and $N,\; N^i$ and $h_{ij}$  are respectively, the lapse function, shift vector
and 3-metric defined on the leaves.
Then, we find that
 \bq
 \lb{1.4a}
 \delta S = \zeta^2 \int{\Big[\left(E^{\mu\nu} - 8\pi G_{\ae} T^{\mu\nu}\right) \delta{g}_{\mu\nu} } 
 + \AE_{\mu}\delta{u}^{\mu}\Big],
 \eq
 where $T^{\mu\nu} \equiv -  \left({2\delta \left(\sqrt{-g} {\cal{L}}_{m}\right)}/{\delta g_{\mu\nu}}\right)/\sqrt{-g}$, and  
 \bqn
 \lb{1.4b}
 E^{\mu\nu} &=& R^{\mu\nu} - \frac{1}{2}g_{\mu\nu}R - T^{\mu\nu}_{\ae},\nb\\
 T_{\ae}^{\alpha\beta}  
 &=& D_{\mu}\Big[u^{(\beta}J^{\alpha) \mu} - J^{\mu(\alpha}u^{\beta)} - J^{(\alpha\beta)}u^{\mu}\Big]\nb\\
&& + c_1\Big[\left(D_{\mu}u^{\alpha}\right)\left(D^{\mu}u^{\beta}\right) - \left(D^{\alpha}u_{\mu}\right)\left(D^{\beta}u^{\mu}\right)\Big]\nb\\
&& + c_4 a^{\alpha}a^{\beta}    - \frac{1}{2}{\cal{L}}_{\ae} g^{\alpha\beta},\nb\\
\AE_{\mu} &=& \frac{\delta {\cal{L}}_{\ae}}{\delta u^{\mu}} =  2\Big(D_{\alpha} J^{\alpha}_{~~~\mu} - c_4 a_{\alpha} D_{\mu}u^{\alpha}\Big),
 \eqn
 with
$J^{\alpha}_{\;\;\;\mu} \equiv M^{\alpha\beta}_{~~~~\mu\nu}D_{\beta}u^{\nu}$ and $
a^{\mu} \equiv u^{\alpha}D_{\alpha}u^{\mu}$.
It should be noted that $T_{\ae}^{\alpha\beta}$ defined above is different from the one given in \cite{BJS} by a term $\lambda_{\ae} u^{\alpha}u^{\beta}$.
The definition of $\AE_{\mu}$ is also different.
 
In the $\ae$-theory, $g_{\mu\nu}$ and $u^{\mu}$ are independent, and the variations of $S$ with respect to them yield, respectively, the 
Einstein-aether equations, $E^{\mu\nu} = 8\pi G_{\ae} T^{\mu\nu}$, and the aether equation $\AE_{\mu} = 0$. However, the hypersurface-orthogonal
condition relates $u^{\mu}$ with the metric components through Eq.(\ref{1.10b}). 
Then, we obtain, 
\bqn
\lb{1.10d}
\delta{u}^{\mu} &=&  \frac{N^i \delta_{i}^{\mu}-\delta_{T}^{\mu}}{N^2}\delta{N}- \frac{\delta_{i}^{\mu}}{N}  \delta{N}^i,\;\;
\frac{\delta{g}_{\mu\nu}}{\delta{N}} = 2N \delta^{T}_{\mu}\delta^{T}_{\nu},\nb\\
\frac{\delta{g}_{\mu\nu}}{\delta{N}^i} &=& -2N_i \delta^{T}_{\mu}\delta^{T}_{\nu} 
- 2h_{ij}\delta^{(T}_{\mu}\delta^{j)}_{\nu},\\
\frac{\delta{g}_{\mu\nu}}{\delta{h}_{ij}} &=& -N^iN^{j} \delta^{T}_{\mu}\delta^{T}_{\nu} - N^{i} \delta^{(T}_{\mu}\delta^{j)}
-N^{j} \delta^{(T}_{\mu}\delta^{i)}
 -  \delta^{(i}_{\mu}\delta^{i)}_{\nu},\nb
\eqn
where $f_{(ij)} \equiv  (f_{ij} + f_{ji})/2$. Substituting the above expressions  into Eq.(\ref{1.4a}), we find that the variations of $S$ with respect to $N,
N^i$ and $h_{ij}$ yield, respectively,  the Hamiltonian, momentum  constraints, and the dynamical equations, given by, 
\bqn
\lb{Hamilton}
&& {\cal{H}}^{\perp} = 8\pi G_{\ae} \rho_{H}, \\
\lb{Momentum}
&& {\cal{H}}_{i} = 8\pi G_{\ae} s_{i}, \\
\lb{Dyn}
&& {\cal{H}}^{ij} = 8\pi G_{\ae} s^{ij},
\eqn
where
\bqn
\lb{equA}
&& {\cal{H}}^{\perp} \equiv E^{\mu\nu}\frac{\delta{g}_{\mu\nu}}{\delta{N}} + \AE_{\mu}\frac{\delta{u}^{\mu}}{\delta{N}},\;\;\;
\rho_{H} \equiv T^{\mu\nu} \frac{\delta{g}_{\mu\nu}}{\delta{N}},\nb\\
&& {\cal{H}}_{i} \equiv E^{\mu\nu}\frac{\delta{g}_{\mu\nu}}{\delta{N^i}} + \AE_{\mu}\frac{\delta{u}^{\mu}}{\delta{N^i}},  \;\;\;
s_i \equiv T^{\mu\nu} \frac{\delta{g}_{\mu\nu}}{\delta{N^i}},\nb\\
&& {\cal{H}}^{ij} \equiv  E^{\mu\nu}\frac{\delta{g}_{\mu\nu}}{\delta{h}_{ij}},\;\; 
s_{ij} \equiv  T^{\mu\nu} \frac{\delta{g}_{\mu\nu}}{\delta{h}_{ij}}.
\eqn
When $E^{\mu\nu} = \AE_{\mu} = 0$,   we find that
\bq
\lb{ss}
{\cal{H}}^{\perp} = {\cal{H}}_{i} = {\cal{H}}^{ij} = 0, \; (E^{\mu\nu} = \AE_{\mu} = 0).
\eq 
That is, if ($g_{\mu\nu}, u^{\mu})$ is a vacuum solution of the hypersurface-orthogonal 
$\ae$-theory, it is also a vacuum solution ($N, N^i, h_{ij}$) of  the IR limit of the non-projectable HL theory, as shown explicitly in
\cite{Jacobson}, although   conversely  it does not hold in general.

 It is also interesting to note that one of the four coupling constants $c_i$ can be eliminated by the field redefinitions \cite{Foster},
\bq
\lb{1.4aa}
g'_{\mu\nu} = g_{\mu\nu} + (\sigma - 1) u_{\mu}u_{\nu}, \;\;\; {u'}^{\mu}  = \frac{1}{\sqrt{\sigma}}u^{\mu},
\eq
for which the action (\ref{1.1}) remains the same in terms of $g'_{\mu\nu}$, after the replacements, $c_{i} \rightarrow c_i'$
\cite{Foster}, where $\sigma$ is a positive constant. On the other hand, it can be shown that 
\bq
\lb{1.11}
\omega^{\mu} \omega_{\mu} =  a^2 - \big(D_{\alpha}u_{\beta}\big)\big(D^{\alpha}u^{\beta}\big) +   \big(D_{\alpha}u_{\beta}\big)\big(D^{\beta}u^{\alpha}\big),
\eq
where $a^2 \equiv a^{\mu}a_{\mu}$. Then, in the hypersurface-orthogonal case, one can always add a term,
\bq
\lb{1.11a}
\Delta{S}_{\ae} = \frac{1}{16\pi G_{\ae}}\int{\sqrt{- {}^{(4)}g} \; d^4x \Big( \alpha \omega^{\mu} \omega_{\mu} \Big)},
\eq
into $S$, which is effectively to shift the coupling constants $c_i$ to ${c}_i''$, where
${c}_{1}'' = c_1 + \alpha,\; {c}_{2}'' = c_2,\; {c}_{3}'' = c_3  -  \alpha,\;
{c}_{4}'' = c_4  - \alpha$.
With the above gauge freedom, one can see that in the hypersurface-orthogonal case, there are essentially only two independent coupling constants.

 \section{Slowly Rotating Spacetimes}

Setting a  spherical body into slow and uniform rotation about an axis, one expects that the metric outside the body will change from its spherically symmetric geometry to a stationary
and axisymmetric configuration. In this regard,  let us  consider the spacetimes, described by the metric, 
\bqn
\lb{3.1}
ds^2 &=& {N}^2(r)dt^2 - \frac{1}{{f}(r)}\Big(dr + {h}(r) dt\Big)^2  - r^2 d\Omega^2 \nb\\
&& - 2r^2\sin^2\theta\omega(r,\theta) dt d\phi,  
\eqn
where we assume that $\big({N}(r), {f}(r), {h}(r)\big)$ denotes the static spherical vacuum  solutions of the  $\ae$-theory without rotation, and ``slowly rotating" means that 
\bq
\lb{3.2}
|\omega| \ll1.  
\eq
It can be shown that for the metric (\ref{3.1})  $\zeta_{(t)} = \partial_{t}$ is still a timelike Killing vector, while  $\zeta_{(\varphi)} = \partial_{\phi}$  a space-like Killing vector with closed orbit.
So, it indeed represents stationary axisymmetric spacetimes.  

In addition,  the  4-velocity defined by Eq.(\ref{1.10b}) is still hypersurface-orthogonal even for the metric (\ref{3.1}).  One might expect that  the aether may also rotate  
with respect to the chosen frame, so that it has a non-vanishing $\phi$ component,  i.e.,
\bq
\lb{3.2a}
u_{\mu} = N\delta_{\mu}^{t} + u_{3}(r, \theta)\delta^{\phi}_{\mu}.
\eq
While in general this is indeed true,   the  hypersurface-orthogonal condition (\ref{1.10}) requires
\bq
\lb{3.2ba}
\frac{u_{3}'}{u_{3}} - \frac{N'}{N} = 0,\;\;\;
u_{3,\phi}  = 0,
\eq
for which we have $u_{3} =  \ell N$, 
where $\ell $ is a constant. However, this equivalent to the coordinate transformation, $t \rightarrow t + \ell \phi$, and the slowly rotating condition requires 
$\ell \simeq {\cal{O}}(\omega) \ll 1$.  Then, this coordinate transformation will leave the metric (\ref{3.1}) unchanged, after a redefinition of $\omega$ \cite{BS}.
Thus, without loss of generality, we consider only the case where $u_{3} = 0$.   

To process further, we  note that 
 the physics is independent of the gauge, given by Eqs.(\ref{1.4aa})
and (\ref{1.11a}). Thus, without loss of generality, in the following  we set
\bq
\lb{1.12}
\hat{c}_3 = - \hat{c}_1,\;\;\; \hat{c}_4 = 0,
\eq
where quantities with hats denote the ones after performing consequentially the two operations given by Eqs.(\ref{1.4aa}) and (\ref{1.11a}).
This is equivalent to the choice 
\bq
\lb{1.13}
\sigma= \frac{1}{1 - c_{+}},\;\;\;
\alpha = \frac{2c_4 + c_{+}c_{-}}{2(1-c_{+})},
\eq
if one first considers the operation (\ref{1.11a}) and then the one (\ref{1.4aa}).  
Our choice (\ref{1.12})  is the same as that given by   Eq.(15) in \cite{Jacobson}. In terms of $c_i$, we have
\bq
\lb{1.13a}
\hat{c}_{1} = \frac{1}{2}\left(2c_4 + c_{+} + 1\right),\;\;\;
\hat{c}_2 = \frac{c_{123}}{1 - c_{+}}.
\eq

It is interesting to note that $c_{14}$ is invariant under both of the operations,  
and is irrelevant with their ordering,
\bq
\lb{1.14}
c_{14} = c_{14}' = {c}_{14}'' = \hat{c}_{14}.
\eq

Then, to the zeroth-order of $\omega$, we obtain the spherical hypersruface-orthogonal Einstein-aether field equations from 
Eqs.(\ref{Hamilton}), (\ref{Momentum}) and (\ref{Dyn}),
\bqn
\lb{equsA}
\bar{\cal{H}}_{3}&=& 8\pi G_{\ae} \bar{\rho}_{H},\\
\lb{equsB}
\bar{\cal{H}}_{i}&=&8\pi G_{\ae} \bar{s}_{i},\\
\lb{equsC}
\bar{\cal{H}}^{ij}&=& 8\pi G_{\ae} \bar{s}^{ij},
\eqn
where quantities with bars denote the spherical seed solutions. 

To the first-order of $\omega$, we find that 
\bq
\lb{3.2dd}
\delta{\cal{H}}^{\perp} = 0, 
\eq
while  the non-vanishing components of $ \delta{\cal{H}}_{i}$ and 
$\delta{\cal{H}}^{ij}$ are given by
\bqn
\lb{3.2d}
\delta{\cal{H}}_{3}&=& -\frac{\sin^2\theta}{N^2}\left[\frac{\left(\sin^3\theta\omega_{,\theta}\right)_{,\theta}}{\sin^3\theta} +
\frac{N\sqrt{f}}{r^2}\left(\frac{r^4\sqrt{f}}{N}\omega'\right)'\right],\nb\\
\delta{\cal{H}}^{13}&=& - \frac{\sqrt{f} h}{2r^4 N}\left(\frac{r^4\sqrt{f}}{N}\omega'\right)',\nb\\
\delta{\cal{H}}^{23}&=& - \frac{\sqrt{f}}{2r^4N}\left[\frac{ r^2h \omega_{,\theta}}{N\sqrt{f}}  + \omega\left(\frac{r^2h}{N\sqrt{f}}\right)'\right]_{,\theta}.
\eqn
 
In the vacuum case, we have 
\bq
\lb{3.2ee}
\delta{\cal{H}}^{\perp}  = \delta{\cal{H}}_{i} =\delta{\cal{H}}^{ij} = 0. 
\eq
When  $h = 0$, from the above expressions we can see that $\delta{\cal{H}}^{ij} = 0$ is satisfied identically, and  the momentum constraint $\delta{\cal{H}}_{i}  = 0$
yields,
\bq
\lb{3.2e}
\frac{\left(\sin^3\theta\omega_{,\theta}\right)_{,\theta}}{\sin^3\theta} +
\frac{N\sqrt{f}}{r^2}\left(\frac{r^4\sqrt{f}}{N}\omega'\right)' = 0, \;\; (h = 0),
\eq
which can be easily solved  by variable separation methods. Since it is not related to black holes, we shall not consider this case any further.
 
When $h \not= 0$,  Eq.(\ref{3.2ee}) yields
\bqn
\lb{3.2ga} 
\left(\sin^3\theta\omega_{,\theta}\right)_{,\theta} &=&0,\\
\lb{3.2gb}
\left(\frac{r^4\sqrt{f}}{N}\omega'\right)' &=& 0,\\
\lb{3.2gc}
\left[\omega' + \left(\frac{h'}{h} -\frac{f'}{2f} - \frac{N'}{N} + \frac{2}{r}\right)\omega\right]_{,\theta} &=& 0.
\eqn
From Eq.(\ref{3.2ga}) we obtain
\bq
\lb{3.2h}
\omega(r,\theta) = \omega_{2}(r)\left[\frac{\cos\theta}{\sin^2\theta} - \ln\left(\tan\frac{\theta}{2}\right)\right] + \omega_{1}(r),
\eq
which is singular at $\theta = 0, \pi$, unless $\omega_{2}(r) = 0$. Then, substituting it into Eq.(\ref{3.2gb}), we obtain
\bq
\lb{3.3}
\omega = - 3J\int{\frac{Ndr}{r^4\sqrt{f}}},
\eq
for which Eq.(\ref{3.2gc}) is satisfied identically.   
Asymptotical flatness condition requires  $f \simeq N^2 \sim 1$, and from  the above we find that
\bq
\lb{3.4}
\omega \simeq \frac{J}{r^3}, \;\;\; (r \gg 1).
\eq
Note that the slowly rotating Kerr black hole has the same limit.   

Therefore, it is concluded that,  {\em for any given spherical vacuum solution $\big(N(r), f(r), h(r)\big)$ of the    $\ae$- theory with the 
hypersurface-orthogonal condition, there always exists a solution $\big(N(r), f(r), h(r), \omega(r, \theta)\big)$, which represents a slowly rotating 
vacuum solution of the  T-theory. When the rotation is switched off, it reduces to the
seed solution  $\big(N(r), f(r), h(r)\big)$}, where $\omega$ is given by Eq.(\ref{3.2ea})  for $h = 0$,  and by Eq.(\ref{3.3}) for $h \not= 0$. 

Note that Eqs.(\ref{3.2ga})-(\ref{3.2gc}) are quite different from Eqs.(11) - (13) given in \cite{BS}. Hence, contrary conclusions regarding to the existence
of slowly rotating black holes were obtained.  

\section{Spherical Static Black Holes in  Einstein-aether Theory} 

Barausse, Jacobson and Sotiriou (BJS)  recently studied spherical static vacuum spacetimes  in the  framework of the $\ae$-theory, and found numerically a class of solutions that represents
black holes. BJS chose to work with the Eddington-Finkelstein coordinates \cite{BJS},
\bq
\lb{2.1}
ds^2 = F(r) dv^2 - 2B(r)dv dr - r^2d\Omega^2,
\eq
in which the four-velocity of the aether is given by
\bq
\lb{2.2}
u^{\alpha}\partial_{\alpha} = A(r)\partial_{v} - \frac{1 - F(r)A^2(r)}{2 A(r) B(r)}\partial_{r}.
\eq 
In the spherical case,  since the aether is always  hypersruface-orthogonal \cite{Jacobson},
 one can introduce the time-like variable $T$, so that  
$u_{\alpha}$ in terms of $T$  takes the form (\ref{1.10a}), from which 
we find that
\bq
\lb{2.5}
dv = \frac{dT}{T_{,v}} + \frac{2A^2B}{1 + A^2F} dr.
\eq
The integrability  condition requires $T_{,vr} = 0$. Without loss of generality, we choose $T_{,v} = 1$, so that Eq.(\ref{2.5}) has the solution,
\bq
\lb{2.6}
v = T + \int^{r}{\left(\frac{2A^2B}{1 + A^2F}\right)dr}.
\eq
Inserting it into Eq.(\ref{2.1}), we find that the metric becomes
\bq
\lb{2.7}
ds^2 =   {N}^2(r)dT^2 - \frac{1}{{f}(r)}\Big(dr + {h}(r) dT\Big)^2  -  r^2 d\Omega^2,
\eq
with 
\bqn
\lb{2.8}
{N}(r) &=& \frac{1+A^2F}{2A},\;\;\;
{f}(r) = \left(\frac{1+A^2F}{2AB}\right)^2,\nb\\
{h}(r) &=& \frac{1- A^4F^2}{4A^2B}.
\eqn
Instead of choosing the gauge(\ref{1.12}), BJS imposed the conditions \cite{BJS,BS},
\bq
\lb{BJSGauge}
\sigma = s_{0}^2 = \frac{c_{123}(2 - c_{14})}{c_{14}(1-c_{+})(2+c_{+} + 3c_2)},\;\;\;
\tilde{c}_{4} = 0. 
\eq
Here  $s_0$ is the speed of the spin-0 mode, so that the spin-0 horizon coincides with the metric horizon
of the redefined metric $g'_{\mu\nu}$.

With the above, BJS were able to show that there exists a class of black hole solutions, characterized by the radii of their horizons,
which is asymptotically flat and free of space-time singularities (except at the origin $r = 0$). 
Using the field redefinitions, one can transform the BJS solutions to the gauge (\ref{1.12}). Since $u_{\mu} = N\delta^{T}_{\mu}$, it is clear
that these field redefinitions only change the lapse function from $N$ to $\sqrt{\sigma} N$, while $N^i$ and $h_{ij}$ remain the same.
As a result,   even in the gauge (\ref{1.12}), the BJS black hole solutions still take the form (\ref{2.7}). Note that these black hole solutions were
rederived in \cite{BSb}.

Taking these black hole solutions as the seeds, from the results presented in the previous section one can see that slowly rotation black 
hole solutions indeed exist in the IR limit of the non-projectable HL theory \cite{BPS}, in contrast to the claim presented in \cite{BS}.

 \section{Conclusions}
 
 In this Comment, we have studied slowly rotating spacetimes, including black holes,  in the non-projectable HL theory \cite{BPS}, by using the 
 equivalence between this version of the HL theory in its IR limit and the $\ae$-theory with hypersurface-orthogonal conditions  \cite{Jacobson}. 
 We have found that slowly rotating black holes indeed exist. This is   in contrast to the results obtained in \cite{BS}.

 It should be noted that the  equivalence between the $\ae$-theory and the IR limit of the non-projectable HL  theory
holds only in the level of action. In particular, the $\ae$-theory still has the general diffeomorphisms as
that of general relativity, while the HL theory has only the foliation-preserving diffeomorphisms. It is exactly because of the former that we are allowed to
make coordinate transformations of the kind $ T = T(v, r)$, which are forbidden by the foliation-preserving diffeomorphisms \cite{Horava}. 

In addition,  it was shown that the static spherical black holes are not stable against non-spherical perturbations \cite{BSb}, although it is not
clear whether or not the high order derivative terms will fix this instability.

 \section*{Acknowlodgements}
 
The author would like to   thank   E. Barausse, T. Jacobson, S. Sibiryakov  and T.  Sotiriou   for valuable discussions and comments. 
This work is supported in part by the DOE  Grant, DE-FG02-10ER41692.


\end{document}